\documentclass[12pt]{article}
\begin{document}
\title{THE LIMITS OF SPECIAL RELATIVITY}
\author{B.G. Sidharth\\
International Institute for Applicable Mathematics \& Information Sciences\\
Hyderabad (India) \& Udine (Italy)\\
B.M. Birla Science Centre, Adarsh Nagar, Hyderabad - 500 063 (India)}
\date{}
\maketitle
\begin{abstract}
The Special Theory of Relativity and the Theory of the Electron have
had an interesting history together. Originally the electron was
studied in a non relativistic context and this opened up the
interesting possibility that lead to the conclusion that the mass of
the electron could be thought of entirely in electromagnetic terms
without introducing inertial considerations. However the application
of Special Relativity lead to several problems, both for an extended
electron and the point electron. These inconsistencies have,
contrary to popular belief not been resolved satisfactorily today,
even within the context of Quantum Theory. Nevertheless these and
subsequent studies bring out the interesting result that Special
Relativity breaks down within the Compton scale or when the Compton
scale is not neglected. This again runs contrary to an uncritical
notion that Special Relativity is valid for point particles.
\end{abstract}
\section{Introduction}
When Einstein proposed his Special Theory of Relativity, there were two ruling paradigms,
which continue to hold sway even today, though not so universally. The first was that of
point elementary particles and the second was that of space time as a differentiable manifold.\\
Little wonder therefore that as the relativistic theory of the
electron developed, there were immediate inconsistencies which were
finally ostensibly resolved only with the intervention of Quantum
Theory. This was because, historically the original concept of the
electron was that of a spherical charge distribution
\cite{rohr,barut,jc}. It is interesting to note that in the
non-relativistic case, it was originally shown that the entire
inertial mass of the electron equalled its electromagnetic mass.
This motivated much work and thought in this interesting direction.
To put it briefly, in non relativistic theory, we get \cite{rohr},
$$\mbox{Kinetic \, energy} =  (\beta /2) \frac{e^2}{Rc^2} v^2,$$
where $R$ is the radius of the electron and $\beta$ is a numerical factor of the order
of $1$. So we could possibly speak of the entire mass of the electron in terms of its
electromagnetic properties.\\
It might be mentioned that it was still possible to think of an electron as a charge
distribution over a spherical shell within the relativistic context, as long as the electron
was at rest or was moving with a uniform velocity. However it was necessary to introduce,
in addition to the electromagnetic force, the Poincare stresses - these were required to
counter balance the repulsive "explosion" of the different parts of the electron.\\
When the electron in a field is accelerated, the above picture no longer holds. We have to introduce the concept of the electron self force which is given by, in the simple case of one dimensional motion,
\begin{equation}
F = \frac{2}{3} \frac{e}{Re^2} \ddot{x} - \frac{2}{3} \ddot{x} +
\gamma \frac{e^2 R}{c^4} \ddot{x} + 0 (R^2)\label{ex}
\end{equation}
where dots denote derivatives with respect to time, and $R$ is the
radius of the spherical electron. More generally (\ref{ex}) becomes
a vector equation. As can be seen from (\ref{ex}), as $R$ the size
of the electron $\to 0$ the first term $\to \infty$ and this is a
major inconsistency. In contrast the second term which contains the
non Newtonian third time derivative remains unaffected while the
third and following terms $\to 0$. It may be mentioned that the
first term (which $\to \infty$) gives the electromagnetic mass of
the
electron while the second term gives the well known Schott term (Cf.ref.\cite{rohr,barut,bgs}).\\
Let us now see how it was possible to rescue the relativistic
electron theory, though at the expense of introducing some
unphysical concepts.
\section{The Advanced and Retarded Fields}
To proceed, from a classical point of view a charge that is
accelerating radiates energy which dampens its motion. This is given
by the well known Maxwell-Lorentz equation, which in units $c = 1$,
and $\tau$ being the proper time, while $\imath = 1,2,3,4,$ is
(Cf.\cite{nar}), is
\begin{equation}
m\frac{d^2x^\imath}{d\tau^2} = eF^{\imath}_{k} \frac{dx^k}{d\tau} + \frac{4e}{3} g_{lk}
\left(\frac{d^3x^\imath}{d\tau^3} \frac{dx^l}{d\tau} - \frac{d^3x^l}{d\tau^3}
\frac{dx^\imath}{d\tau}\right) \frac{dx^k}{d\tau},\label{e1}
\end{equation}
The first term on the right side is the usual external field while the second term is the
damping field which is added ad hoc by the requirement of the energy loss due to radiation.
In 1938 Dirac introduced instead
\begin{equation}
m\frac{d^2x^\imath}{d\tau^2} = e\{F^\imath_k + R^\imath_k\} \frac{dx^k}{d\tau}\label{e2}
\end{equation}
where
\begin{equation}
R^\imath_k \equiv \frac{1}{2} \{F^{\imath}_{k(ret)} - F^{\imath}_{k(adv)}\}\label{e3}
\end{equation}
In (\ref{e3}), $F_{(ret)}$ denotes the retarded field and $F_{(adv)}$ the advanced field.
While the former is the causal field where the influence of a charge at $A$ is felt by a
charge at $B$ at a distance $r$ after a time $t = \frac{r}{c}$, the latter is the advanced
field which acts on $A$ from a future time. In effect what Dirac showed was that the
radiation damping term in (\ref{e1}) or (\ref{e2}) is given by (\ref{e3}) in which an
antisymmetric difference of the advanced and retarded fields is taken. Let us elaborate a
little further.\\
The Maxwell wave equation has two independent solutions, one having
support on the future light cone, this is the retarded solution and
the other having support on the past light cone which has been
called the advanced solution. The retarded solution is selected to
describe the physical situation in conventional theory taking into
account the usual special relativistic concept of causality. This
retarded solution is physically meaningful, as it describes
electromagnetic radiation which travels outward from a given charge
with the speed of light and reaches another point at a later
instant. It has also been called for this reason the causal
solution. On the grounds of this causality, the advanced solution
has been
rejected, except in a few formulations like those of Dirac above, or Feynman and
Wheeler (F-W) to be seen below.\\
In the F-W formulation, the rest of the charges in the universe
react back on the original electron through their advanced waves,
which arrive at the given charge at the same time as the given
charge radiates its electromagnetic waves. More specifically, when
an electron is accelerated at the instant $t$, it interacts with the
other charges at a later time $t' = t + r/c$ where $r$ is the
distance of the other charge--these are the retarded interactions.
However the other charges react back on the original electron
through their advanced waves, which will arrive at the time $t' -
r/c = t$. Effectively, there is instantaneous action at a distance.
It must be mentioned that in the F-W formulation there is no self
force or radiation damping. This is provided instead by the
action of all other charges in the universe on the original charge. There is also no
electromagnetic mass, like the first term on the right side of (\ref{ex}).\\
It must also be mentioned that Dirac's prescription lead to the so
called runaway solutions, with the electron acquiring larger and
larger velocities in the absence of an external
force \cite{hoyle}. This he related to the infinite self energy of the point electron.\\
To elaborate further we use the difference of the advanced and
retarded fields (that is (\ref{e4}) in (\ref{ex}), in the following
manner: We use successively $F_{(ret)}$ and $F_{(adv)}$ in
(\ref{ex}) and take the difference in which case the self force
becomes (Cf.\cite{bgs})
$$F = -\frac{2}{3} \frac{e^2}{c^3} \frac{d}{dt} (\ddot{x}) + 0(R)$$
In the above, the troublesome infinity generating term of (\ref{ex})
is absent, while the third derivative term is retained. On the other
hand this term is required on grounds of conservation of energy, due
to the fact that an accelerated electron radiates energy
(Cf.\cite{bgs2}). Except for the introduction of advanced fields, we
have infinity free results. However, in this formulation too, there
is no electromagnetic mass term, and further, as will be seen below,
we have to extend our considerations to a small neighborhood of the
electron, and not just the point electron itself. To see this in
detail, we observe that the well known Lorentz Dirac equation
(Cf.\cite{rohr}), can be written as
\begin{equation}
ma^\mu (\tau) = \int^{\infty}_{0} K^\mu (\tau + \alpha \tau_0) e^{-\alpha} d\alpha\label{e5}
\end{equation}
where $a^\mu$ is the accelerator and
$$K^\mu (\tau) = F^\mu_{in} + F^\mu_{ext} - \frac{1}{c^2} Rv^\mu ,$$
\begin{equation}
\tau_0 \equiv \frac{2}{3} \frac{e^2}{mc^3} \sim 10^{-23}sec\label{e6}
\end{equation}
and
$$\alpha = \frac{\tau' - \tau}{\tau_0} ,$$
where $\tau$ denotes the time and $R$ is the total radiation rate. Incidentally this is a
demonstration of the non locality in Compton time $\tau_0$,
referred to above.\\
It can be seen that equation (\ref{e5}) differs from the usual equation of Newtonian Mechanics, in that it is non local in time. That is, the acceleration $a^\mu (\tau)$ depends on the force not only at time $\tau$, but at subsequent times also. Let us now try to characterise this non locality. We observe that $\tau_0$  given by equation (\ref{e6}) is the Compton time $\sim 10^{-23}secs$. So equation (\ref{e5}) can be approximated by
\begin{equation}
ma^\mu (\tau) = K^\mu (\tau + \xi\tau_0) \approx K^\mu (\tau)\label{e7}
\end{equation}
Thus as can be seen from (\ref{e7}), the Lorentz-Dirac equation differs from the usual local theory by a term of the order of
\begin{equation}
\frac{2}{3} \frac{e^2}{c^3} \dot{a}^\mu\label{e8}
\end{equation}
the so called Schott term. It is well known that the time component of the Schott term (\ref{e8}) is given by (Cf.ref.\cite{rohr})
$$-\frac{dE}{dt} \approx R \approx \frac{2}{3} \frac{e^2 c}{r^2} \left(\frac{E}{mc^2}\right)^4,$$
where $E$ is the energy of the particle.  Whence integrating over the period of non locality $\sim \tau_0$ the Compton time, we can immediately deduce that $r$ the scale of spatial non locality is given by
$$r \sim c\tau_0,$$
which is of the order of the Compton wavelength.\\
So far as the breakdown of causality is concerned, this takes place
within a period $\sim \tau$, the Compton time as we briefly saw
\cite{rohr,hoyle}. It was at this stage that Wheeler and Feynman
reformulated the above action at a distance formalism in terms of
what has been called their Absorber Theory. In their formulation,
the field that a charge would experience because of its action at a
distance on the other charges of the universe, which in turn would
act back on the original charge is given by
\begin{equation}
re = \frac{2e^2 d}{3dt} (\ddot{x})\label{e4}
\end{equation}
The interesting point is that instead of considering the above force
in (\ref{e4}) at the charge $e$, if we consider the response at an
arbitrary point in its neighborhood as was shown by Feynman and
Wheeler (Cf.ref.\cite{fw}) and, in fact a neighborhood at the
Compton scale, as was argued recently by the author \cite{iaad}, the
field would be precisely the Dirac field given in (\ref{e2}) and
(\ref{e3}).\\
The net force emanating from the charge is thus given by
\begin{equation}
F^{ret} = \frac{1}{2} \left\{F^{ret} + F^{adv}\right\} + \frac{1}{2} \left\{F^{ret} - F^{adv}\right\}\label{e9}
\end{equation}
which is the acceptable causal retarded field. The causal field now
consists of the time symmetric field of the charge together with the
Dirac field, that is the second term in (\ref{e9}), which represents
the response of the rest of the charges. Interestingly in this
formulation we have used a time symmetric field, viz., the first
term of (\ref{e9}) to recover the retarded field with the correct
arrow of time. Feynman and Wheeler stressed that the universe has to
be a perfect absorber or to put it simply, every charged particle in
the universe should respond back to the action on it by the given
charge in our instantaneous action at a distance scenario. In any
case, it was realized that the limits of classical physics are
reached
in the above considerations, at the Compton scale.\\
There are two important inputs which we can see in the above more recent formulation. The first is the action of the rest of the universe at a given charge and the other is minimum spacetime intervals which are of the order of the Compton scale. The minimum spacetime interval removes, firstly the advanced field effects which take place within the Compton time and secondly the infinite self energy of the point electron disappears due to the Compton scale.
\section{Quantum Mechanical Considerations}
The Compton scale comes as a Quantum Mechanical effect, within which we have zitterbewegung
effects and a breakdown of Causal Physics \cite{diracpqm}. Indeed Dirac had noted this aspect
in connection with two difficulties with his electron equation. Firstly the speed of the
electron turns out to be the velocity of light. Secondly the position coordinates become
complex or non Hermitian. His explanation was that in Quantum Theory we cannot go down to
arbitrarily small spacetime intervals, for the Heisenberg Uncertainty Principle would then
imply arbitrarily large momenta and energies. So Quantum Mechanical measurements are an
average over intervals of the order of the Compton scale. Once this is done, we recover
meaningful physics. All this has been studied afresh by the author more recently, in the
context of a non differentiable spacetime and noncommutative geometry \cite{uof}.\\
Weinberg also notices the non physical aspect of the Compton scale \cite{weinberg}. Starting with the usual light cone of Special Relativity and the inversion of the time order of events, he goes on to add, and we quote at a little length and comment upon it, ``Although the relativity of temporal order raises no problems for classical physics, it plays a profound role in quantum theories. The uncertainty principle tells us that when we specify that a particle is at position $x_1$ at time $t_1$, we cannot also define its velocity precisely. In consequence there is a certain chance of a particle getting from $x_1$ to $x_2$ even if $x_1 - x_2$ is space-like, that is, $| x_1 - x_2 | > |x_1^0 - x_2^0|$. To be more precise, the probability of a particle reaching $x_2$ if it starts at $x_1$ is nonnegligible as long as
\begin{equation}
0 \leq (x_1 - x_2)^2 - (x_1^0 - x_2^0)^2 \leq \frac{\hbar^2}{m^2} \cdots\label{ey}
\end{equation}
where $\hbar$ is Planck's constant (divided by $2\pi$) and $m$ is the particle mass. (Such space-time intervals are very small even for elementary particle masses; for instance, if $m$ is the mass of a proton then $\hbar /m = 2 \times 10^{-14}cm$ or in time units
$6 \times 10^{-25}sec$. Recall that in our units $1 sec = 3 \times 10^{10}cm$.) We are thus
faced again with our paradox; if one observer sees a particle emitted at $x_1$, and absorbed
at $x_2$, and if $(x_1 - x_2)^2 - (x_1^0 - x_2^0)^2$ is positive (but less than or
$=\hbar^2 /m^2)$, then a second observer may see the particle absorbed at $x_2$ at a time
$t_2$ before the time $t_1$ it is emitted at $x_1$.\\
``There is only one known way out of this paradox. The second observer must see a particle
emitted at $x_2$ and absorbed at $x_1$. But in general the particle seen by the second
observer will then necessarily be different from that seen by the first."\\
There is another way to view (\ref{ey}). The light cone of special relativity viz., $(x_1 - x_2)^2 - (x^0_1 - x^0_2)^2 = 0$ now gets somewhat distorted because of Quantum Mechanical effects.\\
Let us consider the above in the context of a non zero photon mass. Such a mass $\sim 10^{-65}gms$ was recently deduced by the author, and it is not only consistent with experimental restrictions, but also predicts a new effect viz., a residual cosmic radiation $\sim 10^{-33}eV$, which in fact has been observed \cite{bhtd,mp,newphys,mer,phys}. Such a photon would have a Compton length $\sim 10^{28}cms$, that is the radius of the universe itself.\\
This would then lead to the following scenario: An observer would see a photon leaving a particle $A$ and then reaching another particle $B$, while a different observer would see exactly the opposite for the same event - that is a photon leaves $B$ and travels backward in time to $A$, as in the Weinberg interpretation. This latter gives the advanced potential. The distinction between the advanced and retarded potentials of the old electromagnetic theory thus gets mixed up and we have to consider both the advanced and retarded potentials. We consider this in a little more detail: The advanced and retarded solutions of the wave equation are given by the well known advanced and retarded potentials given by, in the usual notation, the well known expression
$$A^\mu_{ret(adv)} (x) = \frac{1}{c} \int \frac{j^\mu (x')}{|r - r'|} \delta \left(| r - r'| \mp c(t - t')\right) d^4 x'$$
(The retarded part of which leads to the Lienard Wiechart potential of earlier theory).\\
It can be seen in the above that we have the situation described within the Compton wavelength, wherein there are two equivalent descriptions of the same event--a photon leaving the charge $A$ and reaching the charge $B$ or the photon leaving the charge $B$ and reaching the charge $A$. The above expression for the advanced and retarded potentials immediately leads to the advanced and retarded fields  (\ref{e3}) and (\ref{e9}) of the F-W description except that we now have a rationale for this formulation in terms of the photon mass and the photon compton wavelength rather than the perfect absorber ad hoc prescription. In fact there is now an immediate explanation for this of the Instantaneous Action At a Distance Theory alluded to. In this case the usual causal electromagnetic field would be given by half the sum of the advanced and retarded fields. We note that as the photon mass is so small, the usual theory is still a good approximation.\\
To sum up \cite{fw}, the Feynman Wheeler Perfect Absorber Theory
required that every charge should interact instantaneously with
every other charge in the universe, that is that the universe must
be a perfect absorber of all electromagnetic fields emanating from
within. If this condition were satisfied, then the net response of
all charged particles along the future light cone of the given
charge is expressed by an integral that converges. We have argued
that this ad hoc prescription of Feynman and Wheeler as embodied by
the inclusion of the advanced potential is automatically satisfied
if we consider the photon to have a small mass $10^{-65}gms$ as
deduced by the author elsewhere, and which is consistent with the
latest experimental limits-this leading to the effect mentioned by
Weinberg within the Compton wavelength, which is really the
inclusion of the advanced field as well.
\section{The Limits of Special Relativity}
What we have witnessed above is that it is still possible to rescue the classical relativistic theory of the electron, but at the expense of introducing the advanced fields into the physics, fields which have been considered to be unphysical.\\
Another perspective is, as seen above, that there is instantaneous action at a distance, which apparently goes against relativistic causality. But let us now note that in both the Dirac and the Feynman-Wheeler approaches, we are no longer dealing with point particles alone, but rather with a small neighborhood of such a point particle, a neighborhood of a Compton length dimension. Furthermore within the Compton scale, relativistic causality breaks down.\\
We can then reformulate the above considerations in the following manner: The limit of applicability or the limit of validity of the Special Theory of Relativity is the Compton scale. The points within the Compton scale no longer obey Special Relativity and see a non relativistic, instantaneous action at a distance universe.
\section{Discussion}
1. We would like to sum up the foregoing considerations. In
Classical Physics the point electron leads to infinite self energy
via the term $e^2/R$, where $R$ is the radius which is made to tend
to zero. If on the other hand $R$ does not vanish, in other words we
have an extended electron, then we have to introduce non
electromagnetic forces like the Poincare stresses for the stability
of this extended object, though on the positive side this allows the
radiation damping or self force that is required by conservation
laws.\\
Dirac could get rid of the infinity by introducing the difference
between the advanced and retarded potentials: This was the content
of the Lorentz Dirac equation. The new term represents the radiation
damping effect, but we then have to contend with the advanced
potential or equivalently a non locality in time. However this non
locality takes place within the Compton time, within which the
electron attains a luminal velocity.\\
The Lorentz Dirac equation on the other hand had unsatisfactory
features like the derivative of the acceleration, the non locality
in time and the run away solutions, features confined to the Compton scale.\\
The Feynman-Wheeler approach bypasses the infinity and the extended
electron. Moreover the net result is that there is a retarded
potential. But an instantaneous interaction with the rest of the
charges of the universe is required. It is this interaction with the
remaining charges which leads to the point electron's self energy.
Surprisingly however the interaction with the rest of the charges in
the immediate vicinity of the given charge in the Feynman-Wheeler
formula gives us back the Dirac antisymmetric difference with its
non locality within the Compton scale. There is thus a
reconciliation of the Dirac and the Feynman Wheeler approaches, once
we bring into the picture, the Compton scale.\\
In the Feynman-Wheeler approach on the other hand the self force is
dispensed with but at the expense of invoking the instantaneous
interaction of the electron in question with the rest of the charges
in the universe, though even here the Compton scale of the electron
comes into question. Outside this scale, the theory is causal that
is uses only the retarded potential because effectively the advanced
potential gets canceled out as it appears as the sum of the
symmetric and antisymmetric differences. The important point however
is that all this can be explained consistently in the context of the
photon having a non zero mass, consistent with experiment $\sim
10^{-65}gms$.\\
The final conclusion was that in a Classical context a totally
electromagnetic electron is impossible as also the concept of a
point electron. It was believed therefore that the electron was
strictly speaking the subject of Quantum Theory. Nevertheless in
Dirac's relativistic Quantum Electron, we again encounter the
electron with the luminal velocity within the Compton scale,
precisely what was encountered in Classical Theory as well, as noted
above. This again is the feature of a point space time approach. At
this stage a new input was given by Dirac - meaningful physics
required averages over the Compton scale, in which process, the
unphysical zitterbewegung effects were eliminated. Nor has Quantum
Field Theory solved the problem - one has to take recourse to
renormalization, and as pointed out by Rohrlich, one still has a non
electromagnetic electron. In any case, it appears that further
progress would come either from giving up point space time or from
an electron that is extended (or has a sub structure) in some sense
\cite{jc,barut,hoyle,rohr}.\\
2. Nevertheless it is curious to notice that there is some
convergence between the Dirac and the Feynman-Wheeler approaches if
we consider the fact that special relativity, as seen above, does
not hold within the Compton wavelength. This explains the non
locality in time. This justifies the use of the advanced potential
or non locality in time of the Lorentz-Dirac approach or also the
fact that a point inside the Compton wavelength sees a non
relativistic instantaneous action at a distance universe around it -
this is the instantaneous action at a distance of the
Feynman-Wheeler approach. Furthermore, the radiation of photons
emitted by the accelerated electrons (in the Dirac self force) are
meaningful only if they impinge on other charges as in the Field
Theory.\\
We now briefly note the following.\\
3. As Rohrlich \cite{rohr2} observes, "Classical Physics ceases to
be valid at or below the Compton wavelength and this cannot be valid
for a point object."\\
4. The self interaction we encountered above gives rise to radiation
reaction for an extended object which for a point charge appears as
a self acceleration or pre acceleration and an extra inertia, the
electromagnetic mass whose behavior was thought to contradict
special relativity \cite{jc}.\\
5. This apart the contradiction of run away accelerations or a
divergent electromagnetic mass do not apply for an extended
electron.\\
6. If we take special relativity into account, we get the
undesirable factor $4/3$ indicating that part of the mass is not
electromagnetic - the beauty of getting a unified theory for mass is
lost.\\
7. Even in the non relativistic theory, Poincare stresses are
required for the stability of the extended electron.\\
8. On the other hand Fermi and others showed that relativistically
the electromagnetic momentum need not be associated with the
Poynting vector, in which case the undesirable $4/3$ factor does not
arise and there is no need for Poincare stresses.\\
9. In classical relativistic theory, there appeared an impasse. We
could get a special relativistic electron with cohesive forces in an
extended model but at the expense of purely electromagnetic
electron. On the other hand point electrons were not meaningful as
their self energy diverged. Consequently the structure
dependent terms had to be taken seriously.\\
10. We have arrived at the Compton scale from two different
approaches. Classically, there was the electron radius and Quantum
Mechanically the Compton length, both of the same order except for a
factor of the order of the fine structure constant:
$$\hbar / mc \sim \beta \cdot e^2 / mc^2$$
We could consider this to a derivation of the value of the Planck
constant of Quantum Mechanics, in an order of magnitude sense.\\
 11. In any case the
above considerations at the Compton scale lead in recent studies to
a noncommutative geometry and the limit to a point particle no
longer becomes legitimate. This has been discussed
in detail in \cite{uof}.\\

\end{document}